# Electron spill-out effect on third-order optical nonlinearity of metallic nanostructure

Takashi Takeuchi[1,*] and Kazuhiro Yabana[2,**]

[1] Metamaterials Laboratory, RIKEN Cluster for Pioneering Research, Wako, Saitama 351-0198, Japan

[2] Center for Computational Sciences, University of Tsukuba, Tsukuba 305–8577, Japan

[*]Email: takashi.takeuchi.gj@riken.jp; [**]Email: yabana@nucl.ph.tsukuba.ac.jp;

**Abstract** – Over the last three decades, plasmonics using metallic nanostructures has become central to nanophotonics research. Recently, its targets have been extended to nonlinear optical phenomena. In a nonlinear regime, quantum mechanical effects, as exemplified by electron spill-out on the surface of nanostructures, significantly influence the optical response. Quantum hydrodynamic theory (QHT) is a promising basis for the analysis of nonlinear optical responses involving such quantum mechanical effects. Herein, QHT was applied to calculate the third-order optical nonlinearity of a spherical metallic nanostructure. The results demonstrate how electron spill-out strongly affects plasmon resonance and third-order optical nonlinearity.

**Keywords** electron spill-out, nonlinear optical response, quantum hydrodynamic theory, plasmon, metallic nanostructure

# I. INTRODUCTION

Metallic nanostructures that interact with an optical field can generate plasmons, the collective motion of the conduction electrons confined in the nanostructure, which enhances the original field with spatial localization beyond the diffraction limit. Their enhancement- and localization-abilities have attracted extensive attention in the field of nanophotonics [1–4]. Recently, research targets of plasmonic systems using metallic nanostructures have been extended to nonlinear optical applications [5–12]. In a nonlinear regime, the quantum mechanical effects are important. As such quantum effects emerged in nanostructures, there have been the three typical examples: spatial nonlocality [13–20], electron spill-out [21–30], and quantum tunneling [31–41]. Over the last decade, these effects have been theoretically and experimentally confirmed; additionally, some groups have reported on how these effects enhance optical nonlinearity [14, 20, 21, 27, 39, 41], which is essential for the development and improvement in various applications [7–12].

The most straightforward way to numerically predict the quantum mechanical effects is to employ methods that use electronic orbitals; an example is time-dependent density functional theory (TDDFT) in first-principles level or with jellium approximation [42, 43]. However, TDDFT-based numerical calculations are significantly time-consuming because the computational costs are proportional to the product of the number of orbitals and the spatial size of the system. This substantially limits their applicability to structures of several nanometers [34–41]. To overcome the limitation of applied TDDFT, several research groups in the field of plasmonics [13–18, 20–30] have recently introduced semiclassical approaches that employ kinetic energy (KE) functionals that incorporate important quantum effects [44–47]. Such an approach regards the electron dynamics as a fluid that is dependent on local physical quantities, such as the electron density $n(\mathbf{r}, t)$ and electric current density $\mathbf{J}(\mathbf{r}, t)$ or velocity field $\mathbf{v}(\mathbf{r}, t)$. This development eliminates the dependency of the semiclassical approach on electronic orbitals, thus allowing to calculate significantly larger nanostructures at a moderate computational cost (i.e., as compared to TDDFT). Over the last decade, this approach has been employed in the field of plasmonics to describe the quantum mechanical effects of metallic nanostructures. Spatial nonlocality can be described by introducing the Thomas-Fermi (TF) KE functional with an $n$ dependence [13–18, 20]. Thereafter, electron spill-out and quantum tunneling can be incorporated by adopting the von Weizsäcker (vW) KE functional with a $\nabla n$ dependence [21–30]. This is referred to as a TF$\lambda$vW approach, where $\lambda$ is an adjustable and phenomenological parameter that is typically selected to reproduce the electron spill-out predicted by density functional theory (DFT) calculations. Furthermore, a recent study focused on the application of higher-order KE functionals, including the $\nabla^2 n$ contribution, to more precisely describe the electron dynamics [48]. Semiclassical approaches with or beyond TF$\lambda$vW have previously been referred to as quantum hydrodynamic theory (QHT); this appellation is also applied here. As another efficient approach to quantify quantum mechanical effects, time-dependent orbital-free DFT (TDOFDFT) has been proposed and actively studied since the late 1990s [49–54]. TDOFDFT employs the single-orbital Schrödinger equation to express the electron dynamics with KE functionals, thus reducing the computational cost relative to that required for TDDFT. Given that the QHT and TDOFDFT have previously been reported to be inextricably linked [54], in a recent study, QHT was applied to derive a similar single-orbital Schrödinger equation [30].

QHT-based calculations have previously been demonstrated to yield good results when applied to a linear regime to reproduce the quantum mechanical effects of metallic nanostructures [21–29]. However, in the case of nonlinear regimes, applications of QHT remains limited. To the best of our knowledge, only a few studies have focused on using QHT in combination with TF$\lambda$vW (hereafter referred to as QHT-TF$\lambda$vW) to clarify or reproduce nonlinear optical responses [21, 27]. These previous studies demonstrated how electron spill-out affects the interface of a 2D metallic thin film; specifically, they demonstrated that it could induce second-harmonic generation and be directly applied to determine the magnitude of the nonlinear signal. From the perspective of advances in theoretical and practical applications, their research findings are intriguing. However, no reports on the use of QHT elucidate the relationship between electron spill-out and nonlinear optical processes for 3D metallic nanostructures.

In this study, the third-order optical nonlinearity of a metallic nanosphere was investigated by performing QHT-TF$\lambda$vW-based numerical calculations. In particular, how electron spill-out at the surface of the nanosphere contributes to the third-order nonlinearity is clarified here by adjusting the length of the spill-out. Furthermore, the effects of nanosphere size on nonlinearity are also quantified from the perspective of electron spill-out. The results show that the electron spill-out of the nanosphere substantially alters the optical responses in the linear and nonlinear regimes. In particular, it is shown that the third-order optical nonlinearity monotonically increases with increasing electron spill-out volume ratio.

The remainder of this paper is organized as follows. In Sec. II, we briefly introduce QHT and our numerical method to calculate nonlinear optical responses. In Sec. III, the calculated results for nanospheres are presented, where optical nonlinearity originating from the electron spill-out is explored in terms of the electron spill-out length and nanosphere size. Finally, conclusions are presented in Sec. IV.

## II. THEORETICAL FRAMEWORK

In this section, a QHT-TF$\lambda$vW-based numerical approach, recently developed by our group, that discretizes all physical quantities into real-space and real-time grids are briefly reviewed [30]. The investigated system is assumed an isolated metallic nanostructure interacting with an optical field. From here, the QHT is shown in the following equation to describe the electron dynamics by the electric current density $\mathbf{J}(\mathbf{r}, t)$ [26, 30]:

$$\frac{\partial \mathbf{J}}{\partial t} = \frac{ne^2}{m}\mathbf{E} - \frac{e}{m}\mathbf{J} \times \mathbf{B} + \frac{1}{e}\left\{\frac{\mathbf{J}}{n}(\nabla \cdot \mathbf{J}) + (\mathbf{J} \cdot \nabla)\frac{\mathbf{J}}{n}\right\} + \frac{ne}{m}\nabla\left(\frac{\delta E_{\text{XC}}}{\delta n} + \frac{\delta G}{\delta n}\right), \quad (1)$$

where $\mathbf{E}(\mathbf{r}, t)$ and $\mathbf{B}(\mathbf{r}, t)$ are the electric and magnetic fields, respectively, and $n(\mathbf{r}, t)$ is the electron density, which satisfies the equation of continuity, $\partial n/\partial t = (1/e)\nabla \cdot \mathbf{J}$. $\delta E_{\text{XC}}/\delta n = V_{\text{XC}}(\mathbf{r}, t)$ is the exchange-correlation (XC) potential whereas $\delta G/\delta n = V_G(\mathbf{r}, t)$ is the contribution from the KE functional $G$. Eq. (1) can include the XC vector potential term $\mathbf{A}_{\text{XC}}$; however, it is ignored here for simplicity. $m$ and $e$ are the electron mass and elementary charge, respectively. SI units were applied throughout this study.

To describe the spatial nonlocality, electron spill-out, and quantum tunneling, in TF$\lambda$vW-based approaches, $G$ in Eq. (1) is given as follows:

$$G = T_{\text{TF}} + \lambda T_{\text{w}}, \quad (2)$$

where $T_{\text{TF}}$ and $T_{\text{w}}$ denote the TF and vW KE functionals, respectively, and their functional derivatives with respect to $n$ are as follows, with $c_{TF} = (\hbar^2/m)(3/10)(3\pi)^{2/3}$:

$$\frac{\delta T_{\text{TF}}}{\delta n} = \frac{5}{3}c_{\text{TF}}n^{2/3}, \quad (3)$$

$$\frac{\delta T_{\text{w}}}{\delta n} = \frac{\hbar^2}{8m}\left(\frac{\nabla n \cdot \nabla n}{n^2} - 2\frac{\nabla^2 n}{n}\right) = -\frac{\hbar^2}{2m}\frac{\nabla^2 \sqrt{n}}{\sqrt{n}}, \quad (4)$$

where $\hbar$ is the reduced Planck's constant. Equation (2) includes an adjustable phenomenological parameter $\lambda$ as the weight of the vW term, the value of which determines the length of the electron spill-out; it is generally selected to range from $1/9$ to $1$ [47]. Intriguingly, this QHT-TF$\lambda$vW approach can be directly derived from a fully quantum mechanical approach, TDDFT, with $\lambda = 1$ and an additional term consisting of electronic orbitals that also contributes to electron spill-out [26]. The next section describes how the length of the electron spill-out can be controlled by applying different values of $\lambda$ and details the investigation into its influence on the nonlinear optical response of metallic nanostructures.

As mentioned before, previous studies on linear regime have already shown QHT-TF$\lambda$vW-based calculations to yield results that effectively clarify how electron spill-out affects the optical response, particularly in terms of the plasmon resonance. However, it is considerably difficult to develop computational codes that can stably execute Eqs. (1)–(4) under nonlinear conditions. The reason for the difficulty is related to the spatiotemporal changes in the electron density that are governed by $\partial n/\partial t = (1/e)\nabla \cdot \mathbf{J}$ that, in a linear regime, can be handled without problem by linearizing the equation. Equations (3) and (4), which respectively include $n^{1/3}$ and $1/n$ components, presuppose $n$ to be a positive definite quantity; thus, the numerical calculation process will immediately break down if the update of $n$ induces a numerical error that results in even a single spatial and/or temporal point becoming $n \leq 0$. To mitigate this problem, a new numerical scheme was developed to stably analyze the nonlinear optical response of metallic nanostructures using a TF$\lambda$vW approach. The numerical scheme utilized to rewrite the basic equations for the QHT-TF$\lambda$vW system (i.e., Eqs. (1)–(4)) to the following effective Schrödinger equation (ESE) [30] with $\lambda = \xi^2$:

$$i\hbar\xi \frac{\partial \Psi}{\partial t} = \left[ \frac{(-i\hbar\xi\nabla + e\mathbf{A})^2}{2m} - e\phi + \frac{\delta E_{\text{XC}}}{\delta n} + \frac{\delta T_{\text{TF}}}{\delta n} \right]\Psi, \quad (5)$$

where $\mathbf{A}(\mathbf{r},t)$ and $\phi(\mathbf{r},t)$ respectively denote the vector and scalar potentials related to the electromagnetic fields as $\mathbf{E} = -\partial \mathbf{A}/\partial t - \nabla\phi$ and $\mathbf{B} = \nabla \times \mathbf{A}$. $\Psi(\mathbf{r},t)$ is the effective wave function, which produces the electron density $n$ and electric current density $\mathbf{J}$ as follows:

$$n = |\Psi|^2, \quad (6)$$

$$\mathbf{J} = -\frac{e}{m} \text{Re}[\Psi^*(-i\hbar\xi\nabla + e\mathbf{A})\Psi]. \quad (7)$$

Equations (5)–(7) can be directly derived from Eqs. (1)–(4) with no assumptions, as they are equivalent. (Details are provided in our previous paper [30]). Nevertheless, it can be easily understood that the numerical stability of the ESE-based calculations is better than that of the original QHT-based calculations using Eqs. (1)–(4). Particularly, (i) vW potential of Eq. (4) does not appear in Eq. (5), while it appears in TDOFDFT [49-54]. The vW term is treated through a kinetic operator with $\hbar \to \hbar\xi$, and thus Eq. (5) does not include a singular point with respect to $1/n$; (ii) Eq. (6) guarantees $n$ to be a positive definite quantity $n \geq 0$. This is the reason why the numerical calculations presented as Eqs. (5)–(7) are stably executable for the nonlinear optical phenomena described in a QHT-TF$\lambda$vW system.

The electromagnetic potentials that appear in Eqs. (5) and (7), i.e., $\mathbf{A}$ and $\phi$, can be updated by applying the Lorentz gauge condition to Maxwell's equations, as follows:

$$\frac{\partial \mathbf{E}}{\partial t} = \frac{1}{\epsilon_0 \mu_0} \nabla \times \mathbf{B} - \frac{1}{\epsilon_0}\mathbf{J}, \quad (8)$$

$$\frac{\partial \mathbf{B}}{\partial t} = -\nabla \times \mathbf{E}, \quad (9)$$

$$\frac{\partial \mathbf{A}}{\partial t} = -\mathbf{E} - \nabla \phi, \tag{10}$$

$$\frac{\partial \phi}{\partial t} = -\frac{1}{\epsilon_0 \mu_0} \nabla \cdot \mathbf{A}, \tag{11}$$

In the developed ESE-based numerical scheme, Eqs. (5)–(7) and (8)–(11) are concurrently solved in discretized real-space and real-time grids; details on their implementation are available in Ref. [30].

### III. NUMERICAL RESULTS

This section reveals how the effects of electron spill-out contribute to the nonlinear optical response of a metallic nanostructure in QHT-TF$\lambda$vW-based numerical calculations. The target system is displayed in Fig. 1(a), which shows a metallic nanosphere with a diameter $a$ that is subjected to an applied incident electromagnetic field formed by $\mathbf{E}^{(i)}$ and $\mathbf{B}^{(i)}$. Generally, QHT describes metallic nanostructures through the use of a jellium model (JM) that replaces an original atomistic structure of material to a uniform positive background density $n_{\mathrm{JM}}^{(+)}(\mathbf{r})$ [55]. For the nanosphere applied in this study, $n_{\mathrm{JM}}^{(+)}(\mathbf{r})$ is given by the following expression, with $n_s = \left((4\pi) r_s^3/3\right)^{-1}$:

$$n_{\mathrm{JM}}^{(+)} = \begin{cases} n_s & \text{if } |\mathbf{r}| \leq a, \\ 0 & \text{if } |\mathbf{r}| > a, \end{cases} \tag{12}$$

where $r_s$ is the Wigner-Seitz radius of the medium and has been set to $r_s = 3.99$ Bohr to correspond to Na metal. The application of the JM constitutes a considerable simplification that serves to reduce computational costs. However, by linking to TDDFT or other related numerical schemes including QHT, the JM has so far succeeded in reproducing experimental results, which contain typical quantum mechanical effects that emerged in metallic nanostructures, such as spatial nonlocality, electron spill-out, and tunneling [27, 31–41]. The XC potential $\delta E_{XC}/\delta n$ in Eqs. (1) and (5) can be calculated by adopting adiabatic and local density approximations [56]. All calculations were performed using SALMON, which is an open-source code developed by our group [57].

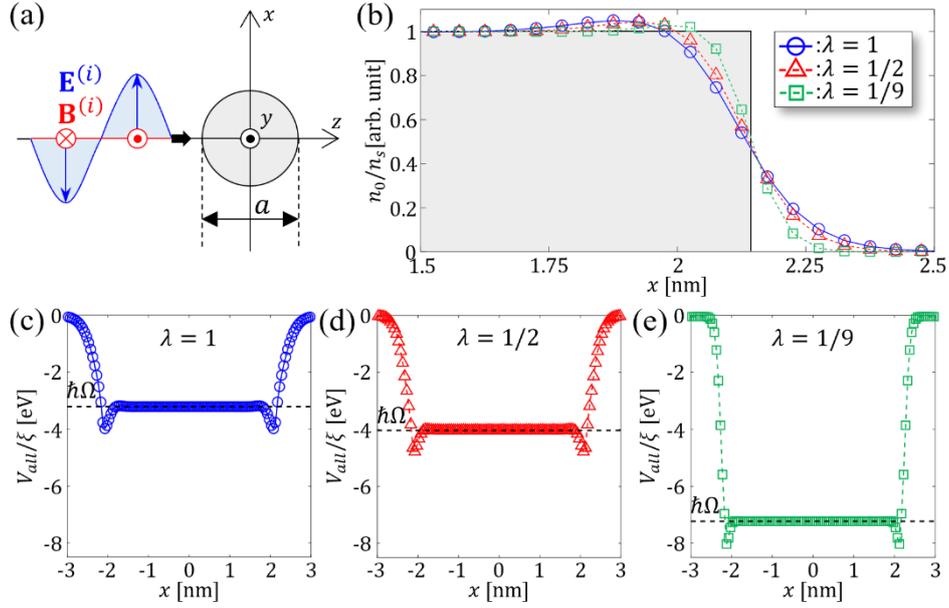

FIG. 1. (a) Target system consisting of the Na nanosphere with a diameter $a$ and the incident light electromagnetic field represented as $\mathbf{E}^{(i)}$ and $\mathbf{B}^{(i)}$, which propagate in the $z$ direction. The center of mass of the nanosphere is located at the origin. Note that $a = 4.3$ nm in (b)–(e). (b) Spatial distribution of initial electron density $n_0$ along the $x$ axis near the surface of the nanosphere. The blue solid line with circles, red dashed line with triangles, and green dashed line with squares respectively show the $n_0$ results for $\lambda = 1$, $1/2$, and $1/9$. The gray shaded area indicates the positive background density $n_{JM}^{(+)}$. Their amplitudes have been normalized with respect to $n_s$. (c)–(e) Spatial distribution of the potential $V_{all}/\xi$ described in Eq. (13), where $\lambda = 1$, $1/2$, and $1/9$, respectively. The horizontal dashed black lines indicate the energy eigenvalue $\hbar\Omega$.

## A. $\lambda$ dependence

Here, the dependence of $\lambda$ is explored by varying the parameter between $1/9$ and $1$, which is the range generally applied to treat the vW correction and determine the length of the electron spill-out [47]. The diameter of the nanosphere, $a$, was set at $4.3$ nm, which corresponds to an electron number of $N_e = 1074$ for Na metal. Although the length of the electron spill-out was adjusted by varying $\lambda$ in this calculation, for actual experiment, such adjustments cannot be made for isolated nanoparticles. However, the electron spill-out can be changed by introducing a coating material that surrounds the nanostructure, thereby forming a core-shell nanostructure [27]. When a small $\lambda$ was employed, there was rapid electron spill-out decay on the order of subnanometers, closing to the hard-wall boundary assumed in classical electromagnetism theory. To reproduce this rapid decay in numerical calculations, we employed particularly fine spatial grid spacings, i.e., $\Delta x = \Delta y = \Delta z = 0.05$ nm, under the conditions of a computational domain set to be a cube with side lengths of 14.6 nm, which is sufficiently larger than $a$.

In QHT, the initial electron density of a metallic nanostructure $n_0(\mathbf{r})$ must be known before the optical response can be calculated. Although several candidates for $n_0(\mathbf{r})$ have been utilized in linear response calculations, including one that was calculated based on DFT or model density [25, 48], the equilibrium density should be used here to investigate nonlinear responses; this density was determined by applying the following condition to Eq. (1): the electric current density $\mathbf{J}$ at $n_0(\mathbf{r})$ completely vanishes anywhere in the spatial grids. In the case of the ESE, the equilibrium density is obtained by substituting the ground state wave function $\Psi_0(\mathbf{r}, t) = \sqrt{n_0(\mathbf{r})} \exp(-i\Omega t)$ in Eq. (5) with $\lambda = \xi^2$ as follows:

$$\hbar\Omega\sqrt{n_0} = \left[-\frac{\hbar^2}{2m}\xi\nabla^2 + \frac{V_{all}}{\xi}\right]\sqrt{n_0}, \tag{13}$$

$$V_{all} = -e\phi_0 + \frac{\delta E_{XC}}{\delta n} + \frac{\delta T_{TF}}{\delta n}, \tag{14}$$

where $\hbar\Omega$ corresponds to the energy eigenvalue of $\Psi_0$. Then, $\phi_0(\mathbf{r})$ in Eq. (14) is the static scalar potential that satisfies the following Poisson's equation that considers $n_{JM}^{(+)}$

and $n_0$:

$$\nabla^2 \phi_0 = -\frac{e\left(n_{\text{JM}}^{(+)} - n_0\right)}{\epsilon_0}. \tag{15}$$

We employed the multipolar expansion to determine the boundary value of $\phi_0$. $\sqrt{n_0}$ was obtained by applying an imaginary time-propagation method [30, 58].

Figure 1(b) shows the $n_0$ results calculated for $\lambda = 1$ (the solid blue line with circles), $1/2$ (the dashed red line with triangles), and $1/9$ (the dashed green line with squares); each $n_0$ was shown for the $x$ axis near the surface of the nanosphere and normalized with respect to $n_s$. As the figure indicates, $\lambda$ is a direct determinant of the electron density distribution; particularly, the electron spill-out decreases with $\lambda$. This behavior with respect to the density tail has been well documented by previous authors who have conducted related QHT studies [21–29]. Thus, to further analyze these results, we plotted the spatial distribution of $V_{all}/\xi$ that confines $n_0$, as described by Eq. (13). Figures 1(c)–(e) show the results for $\lambda = 1$, $1/2$, and $1/9$, respectively; the horizontal dashed black line in each figure indicates the energy eigenvalue $\hbar\Omega$. As can be seen, the same three general trends were observed for all values of $\lambda$: (i) a flat line at $V_{all}/\xi \approx \hbar\Omega$ inside the sphere ($|x| < 1.7$ nm); (ii) two spikes extending in a negative direction at the points along the $x$ axis near the surface of the sphere ($1.7 \leq |x| \leq 2.2$ nm); and (iii) a sharp increase toward $V_{all}/\xi = 0$ outside of the sphere ($|x| > 2.2$ nm). For all values of $\lambda$, $\hbar\Omega$ was calculated to be close to the bottom of $V_{all}/\xi$, the value of which decreased with $\lambda$. In the asymptotic region where $V_{all}$ is ignorably small, the density behaves as $n_0 \sim e^{-2\kappa r}$ with $\kappa = \sqrt{-2m\Omega/\hbar\xi}$. Since the magnitude of the eigenvalue $|\hbar\Omega|$ increases as $\xi = \sqrt{\lambda}$ decreases, the length of electron spill-out decreases as $\lambda$ decreases, as shown in Fig. 1(b).

Hereafter, the focus will be on real-time calculations performed using Eqs. (5)–(11). The temporal interval $\Delta t$ was set as $9.628 \times 10^{-5}$ fs to satisfy the Courant-Friedrichs-Lewy condition for the applied spatial grid spacing conditions, i.e., $\Delta x = \Delta y = \Delta z = 0.05$ nm [59]. We employed the perfect matched layer as the electromagnetic absorbing boundary condition.

Before investigating the nonlinear optical response, the linear properties of the metallic nanospheres were calculated. Generally, a loss mechanism must be incorporated into QHT-based numerical calculations to mimic the damping effect of plasmons. For example, in previous QHT studies that entailed the use of Eq. (1), a friction term was added to the equation with the phenomenological damping rate parameter $\gamma$ [21–29]; however, such a friction term cannot be easily introduced into the ESE (Eq. (5)) for the QHT-based calculations in this study. Thus, in our previous study [30], we proposed another way to replicate the loss mechanism with the ESE by introducing the following conductive current density $\mathbf{J}_C(\mathbf{r}, t)$:

$$\mathbf{J}_C(\mathbf{r}, t) = \sigma g(\mathbf{r})[\mathbf{E}(\mathbf{r}, t) - \mathbf{E}_{gs}(\mathbf{r})], \tag{16}$$

where $\sigma$ is a phenomenological conductivity parameter, and the second term in square brackets $\mathbf{E}_{gs}(\mathbf{r})$ is the electric field in the ground state and has been added to maintain $\mathbf{J}_C = 0$ at the ground state prior to light irradiation. $g(\mathbf{r})$ specifies the spatial distribution of $\mathbf{J}_C$ and is modeled as

$$g(\mathbf{r}) = \frac{n_0(\mathbf{r})}{n_s}. \tag{17}$$

The total electric current density in Eq. (8) can be modified as $\mathbf{J}(\mathbf{r}, t) = \mathbf{J}_Q(\mathbf{r}, t) + \mathbf{J}_C(\mathbf{r}, t)$, where $\mathbf{J}_Q$ denotes the original value defined in Eq. (7). In this study, $\sigma =$ $5.05 \times 10^3$ S/m ($1.10 \times 10^{-3}$ a.u.) was applied, the value of which could reproduce the linear and nonlinear optical resonances calculated by TDDFT for the same JM nanosphere [30].

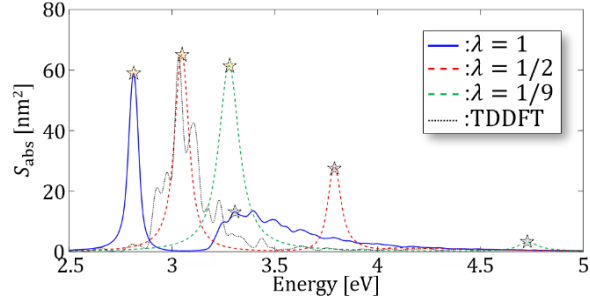

FIG. 2. Spectral distribution of the linear optical absorption cross-section $S_{abs}$ of the nanosphere ($a = 4.3$ nm). The solid blue and dashed red and green lines show $S_{abs}$ for $\lambda = 1, 1/2,$ and $1/9$ using QHT, respectively, whereas the dotted black line is that calculated by TDDFT. The yellow shaded stars found in QHT spectra indicate the peak position of the plasmon resonance, whereas the gray shaded stars denote the Bennett state.

Figure 2 displays the spectral distribution results for the linear optical absorption cross-section $S_{abs}$ of the metallic nanosphere with $a = 4.3$ nm; the calculation process to obtain $S_{abs}$ through the ESE is described in detail in our previous paper [30]. As shown in Fig. 1(b)–(e), $S_{abs}$ was calculated for $\lambda = 1, 1/2,$ and $1/9$ using QHT, and the results are respectively presented as the solid blue, dashed red, and dashed green lines. Additionally, we plotted $S_{abs}$ calculated by TDDFT as a reference that is presented as the dotted black line. In our previous study [30], we mentioned that QHT with $\lambda = 1/2$ could well mimic the electron spill-out calculated in DFT, therefore reproducing optical responses almost identical to those for TDDFT in both the linear and nonlinear regimes. In the present study, we explore how the electron spill-out controlled by $\lambda$ in QHT calculation differentiates their optical responses. Each peak of QHT results appeared in the lower-energy region corresponds to the plasmon resonance, the position of which is marked by a yellow shaded star. As can be ascertained from the QHT results, the stars were blueshifted as $\lambda$ decreases. This blueshift reflects the impact of the electron spill-out on the linear optical response, as has been confirmed in previous reports on QHT studies [21–29]. The gray shaded stars in the higher-energy region indicate the spectral position of the Bennett state, which is induced by the nonuniformity of the electron density, to which the spill-out is a contributing factor, and is known to be overestimated in QHT-based calculations as can be seen in comparison with $\lambda = 1/2$ and TDDFT [25, 38, 48, 60]. The spectral area of the Bennett state, which is equivalent to the number of electrons that are photo-excited through the state, decreases for smaller values of $\lambda$. This can be explained by the change in $n_0$ with respect to $\lambda$, as indicated in Fig. 1(b), which shows that there was less electron spill-out for smaller values of $\lambda$, such as $\lambda = 1/9$. This phenomenon indicates that $n_0$ in the nanosphere tends to be uniform with a hard-wall boundary as is used in classical electromagnetism. The Bennett states observed at $\lambda = 1/2$ and $1/9$ had a single peak; however, the results for $\lambda = 1$ show several fragmented peaks at energies higher than 3.2 eV. Recently, Ciracì et. al. mentioned that the spectral structure of the Bennett state is not only often overestimated but also strongly dependent on the size of the computational domain [48]. This is consistent with the results presented in Fig. 1(c)–(e), which show that the magnitudes of the energy differences between $\hbar\Omega$ and 0 eV for $\lambda = 1/2$ and $1/9$ were larger than their respective Bennett state energies (Fig. 2), whereas that for $\lambda = 1$ was smaller than its Bennett state. This means that the Bennett state appears, in the ESE system, as the bound excited state below the threshold for $\lambda = 1/2$ and $1/9$, and as the continuum excited state above the threshold for $\lambda = 1$. Therefore, in consequence, artificial quantum states, the energy levels of which are characterized by the inverse of computational domain size, interferes with the Bennett state for $\lambda = 1$, producing the many fragmented peaks that can be observed in Fig. 2 (blue line). To obtain a fully convergent spectrum under the conditions of $\lambda = 1$, a wider computational domain is required; however, this was not applied herein because the focus was solely on the plasmon resonance with electron spill-out not the Bennett state overestimated in QHT.

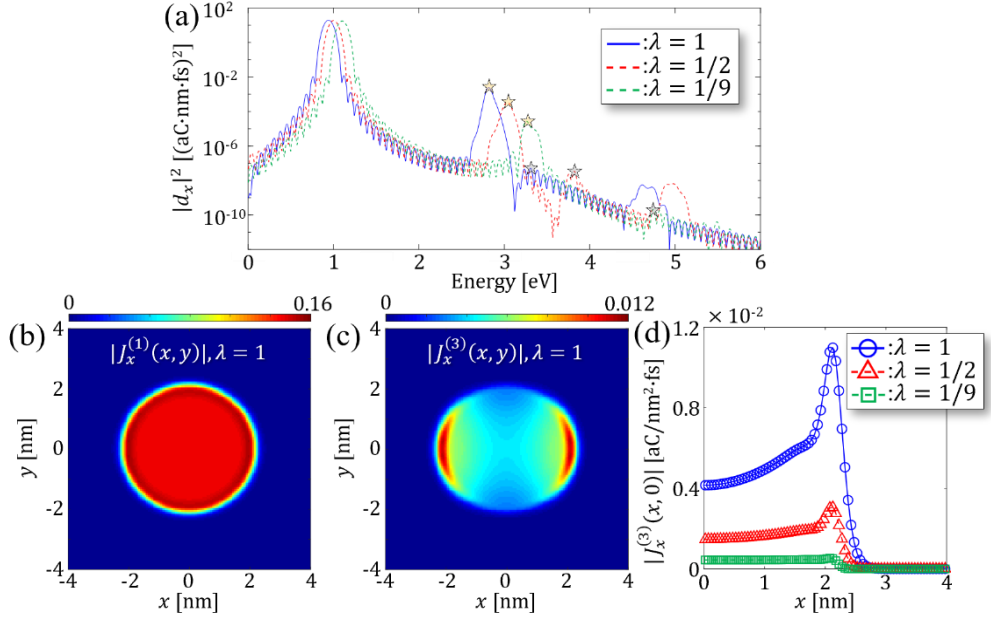

FIG. 3. (a) Power spectra for the $x$ component of the total dipole moment $d_x$ for a nanosphere with $a = 4.3$ nm. The solid blue and dashed red and green lines correspond to $\lambda = 1, 1/2$, and $1/9$, respectively. The fundamental frequencies of the incident pulse $\omega_i$ were set to be 0.94, 1.00, and 1.10 eV for $\lambda = 1, 1/2$, and $1/9$, respectively. The yellow shaded stars indicate the peak position of the plasmon resonance, whereas the gray shaded stars denote the Bennett state. (b)–(c) Spatial distributions of $|J_x^{(k)}(x,y)|$ for $k = 1$ and 3, respectively (see Eq. (21)), as calculated for the $\lambda = 1$ condition. (d) Spatial distributions of $|J_x^{(3)}(x, y = 0)|$ for the $\lambda = 1, 1/2$, and $1/9$ conditions; the results are respectively presented as a solid blue line with circles, dashed red line with triangles, and dashed green line with squares.

Let us now explore the nonlinear optical response of the metallic nanospheres with $a = 4.3$ nm. For the incident light illustrated in Fig. 1(a), the following pulsed electric field $\mathbf{E}^{(i)}$ was employed:

$$\mathbf{E}^{(i)}(t) = F \cos^2\left[\frac{\pi}{T}\left(t - \frac{T}{2}\right)\right] \sin(\omega_i t)\, \hat{\mathbf{x}} \quad (0 < t < T), \tag{18}$$

where $F$, $T$, and $\omega_i$ are the pulse parameters specifying the amplitude, duration, and fundamental frequency, respectively, and $\hat{\mathbf{x}}$ denotes the unit vector along the $x$ axis. Hereafter, the values of $F$ and $T$ have been fixed at 274 MV/m (corresponding to $5.34 \times 10^{-4}$ a.u. and $I = 10^{10}$ W/cm$^2$) and 55 fs, respectively.

Figure 3(a) shows the power spectra for the $x$ component of the total dipole moment $d_x$, which is defined by the following expression:

$$d_x(t) = \hat{\mathbf{x}} \cdot \int_0^t dt \left[ \int \int \int dxdydz \{\mathbf{J}_Q(\mathbf{r},t) + \mathbf{J}_C(\mathbf{r},t)\} \right]. \tag{19}$$

The following window function $w(t)$ is also used to eliminate the spurious oscillations that originate from the finite period of the Fourier transformation:

$$w(t) = 1 - \frac{1}{3}\left(\frac{t}{T_{\max}}\right)^2 + 2\left(\frac{t}{T_{\max}}\right)^3, \tag{20}$$

where $T_{\max}$ is the duration of the real-time computation and has been fixed at $2T$ in this study. The calculated $|d_x(\omega)|^2$ results for $\lambda = 1$, $1/2$, and $1/9$ are shown in Fig. 3(a). The fundamental frequencies of the incident pulse $\omega_i$ were set at 0.94, 1.00, and 1.10 eV for the $\lambda = 1$, $1/2$, and $1/9$ conditions, respectively. These $\omega_i$ values approximately satisfy one-third of each plasmon resonance energy; thus, their third-harmonic generation was plasmonically enhanced. In Fig. 3(a), it can be seen that the spectral positions of the plasmon resonance marked by the yellow shaded stars, which overlap with their third-harmonic generation, emerged near 3 eV. It is also intriguing that the third nonlinear signals increased with increasing $\lambda$, whereas the magnitudes of the fundamental signals around 1 eV were nearly the same. Other peaks have been classified as those enhanced by the Bennett state and denoted by gray shaded stars; the fifth-harmonic generation occurred at approximately 5 eV.

To explore the mechanism that determines the magnitude of the third-harmonic generation, the $x$ component of the $k$-th order electric current density distribution $J_x^{(k)}(x, y)$ was evaluated; it is defined as

$$J_x^{(k)}(x, y) = \hat{\mathbf{x}} \cdot \int_0^{T_{\max}} dt \left[ w(t) \{ \mathbf{J}_Q(x, y, z = 0, t) + \mathbf{J}_C(x, y, z = 0, t) \} e^{ik\omega_i t} \right]. \quad (21)$$

Figure 3(b)–(c) respectively show the results for $k = 1$ and $3$ for the $\lambda = 1$ condition. As shown in Fig. 3(b), the linear component, which corresponds to $k = 1$, has an almost uniform current distribution. In contrast, as shown in Fig. 3(c), in the case of the third-order nonlinear component (i.e., $k = 3$), the current is more enhanced at the edges of the nanosphere, the locations of which reflect electron spill-out, as indicated by the result shown in Fig. 1(b). To clearly elucidate the respective relationships between the nonlinear current and electron spill-out, $|J_x^{(3)}(x, y = 0)|$ was plotted for $\lambda = 1, 1/2$, and $1/9$; the results are shown in Fig. 3(c). Each maximum $|J_x^{(3)}(x, y = 0)|$ value can be observed corresponding to a position near the surface of the nanosphere; furthermore, the current in whole spatial region is more than linear increase with respect to $\lambda$. This clearly indicates that electron spill-out is important for the third-order optical nonlinearity.

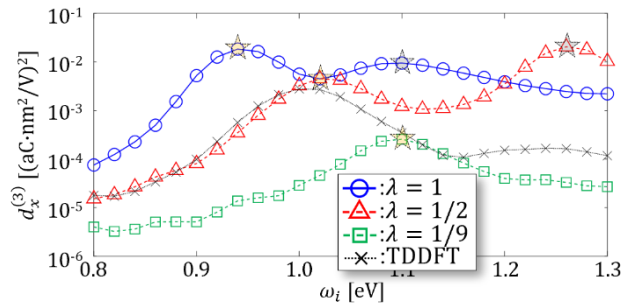

FIG. 4. $\omega_i$ dependence of the third-order optical nonlinearity $d_x^{(3)}$, as defined in Eq. (22) for the nanosphere with $a = 4.3$ nm. The solid blue line with circles, dashed red line with triangles, and dashed green line with squares respectively denote the results for the $\lambda = 1$, $1/2$, and $1/9$ conditions using QHT, whereas the dotted black line with crosses is that calculated by TDDFT. The yellow shaded stars found in QHT results denote the $\omega_i$ values that approximately satisfy one-third of the plasmon resonance energy, whereas the gray shaded stars denote the Bennett state.

The results in Fig. 3 are presented with a focus on the one $\omega_i$ value that approximately satisfies one-third of the plasmon resonance energy. Thus, to investigate the dependence of $\omega_i$ for nonlinear responses, the following third-order optical nonlinearity along the $x$ axis, i.e., $d_x^{(3)}(\omega_i)$, has been defined as follows:

$$d_x^{(3)}(\omega_i) = \frac{\int_{2.5\omega_i}^{3.5\omega_i} d\omega \, [|d_x(\omega)|^2]}{\int_0^\infty d\omega \left[\left|E_x^{(i)}(\omega)\right|^2\right]}, \tag{22}$$

where $E_x^{(i)}(\omega)$ and $d_x(\omega)$ are obtained by applying a Fourier transformation to Eqs. (18) and (19) with the window function $w(t)$ defined in Eq. (20). Here, $d_x^{(3)}(\omega_i)$ denotes the magnitude of the third-harmonic generation for an incident pulse with $\omega_i$. Figure 4 shows the $d_x^{(3)}$ results for $\lambda = 1, \ 1/2,$ and $1/9$ using QHT. To compare the archived nonlinearities, here again, we plotted $d_x^{(3)}(\omega_i)$ calculated by TDDFT as a reference. The yellow shaded stars associated with QHT results in the figure denote the $\omega_i$ values that approximately satisfy one-third of the plasmon resonance energy, as discussed in Fig. 3(a), whereas the gray shaded stars denote the Bennett state. Each yellow shaded star indicates a peak, the value of which shows more than linear increase with respect to $\lambda$, as has been illustrated in Fig. 3(d). These results again confirm that the plasmonically enhanced nonlinear process is strongly influenced by the electron spill-out. The calculation with $\lambda = 1/2$ that reproduces the electron spill-out in the TDDFT calculation [30] also shows accurately the third-order nonlinear response in the plasmonically enhanced region. The gray shaded stars that mark the other peaks in the higher $\omega_i$ region of the results for $\lambda = 1$ and $1/2$ are attributable to the Bennett state, which, as previously mentioned, is overestimated in QHT-based calculations, and therefore should not be observed in actual measurements. Actually, there appears no such peak but a gentle shoulder in the TDDFT result in comparison with QHT calculation using $\lambda = 1/2$. Note that there is no gray shaded star for the $\lambda = 1/9$ condition because the corresponding value for the Bennett state was found to be outside the bounds of this plot, and because its influence was found to be much smaller than that for the other $\lambda$ conditions, as illustrated in Fig. 2.

**B. Size dependence**

In the previous subsection, the nonlinear optical property of a metallic nanosphere with a diameter $a = 4.3$ nm was explored from the perspective of $\lambda$ dependence appeared in Eq. (2), which directly controlled the length of the electron spill-out. This subsection presents an investigation into the size dependence of the metallic nanospheres in the nonlinear optical response. Here, the value of $\lambda$ is fixed at $1/2$, which has been demonstrated to yield an accurate reproduction of the electron spill-out results calculated by DFT-based calculations with the JM for a metallic nanosphere without a coating material [30]. This $\lambda$ value allows for the use of grid spacing conditions that are less precise than those used in the previous subsection; this is because fine grids were essential to reproduce rapidly decayed electron spill-out under the condition of $\lambda = 1/9$. Thus, the spatial grid spacing applied to investigate the size dependence was set at $\Delta x = \Delta y = \Delta z = 0.1$ nm. Additionally, the application of the Courant-Friedrichs-Lewy condition also allows for the extension of the temporal interval to $\Delta t = 1.925 \times 10^{-4}$ fs. The computational domain size was set at $3a$. The other parameters and conditions were the same as those described in the previous subsection.

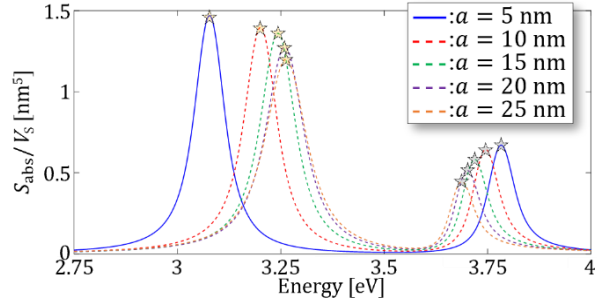

FIG. 5. Spectral distributions of the linear optical absorption cross-section $S_{\text{abs}}$ of nanospheres with $\lambda = 1/2$ and $a = 5$ (solid blue line), $10$ (dashed red line), $15$ (dashed green line), $20$ (dashed purple line), and $25$ (dashed orange line) nm. The vertical axis has been normalized with respect to $V_S$, i.e., the volume of the nanospheres. The yellow shaded stars indicate the peak positions of the plasmon resonance, and the gray shaded stars denote the Bennett state.

Here again, let us begin by investigating the size dependence with respect to the linear optical response. Figure 5 shows the spectral distribution results for the linear optical absorption cross-section $S_{\text{abs}}$ of metallic nanospheres with $a = 5$, $10$, $15$, $20$, and $25$ nm. The electron number $N_e$ contained in these spheres approximately ranged from $1700$ to $208000$. Equations (13)–(15) were applied to determine the corresponding initial electron density $n_0$. For those nanospheres with sufficiently large size, the electron spill-out keeps the same length because $\lambda$ is fixed as $1/2$. Because the spectral area of $S_{\text{abs}}$ is proportional to $N_e = n_s V_S$, where $V_S$ denotes the volume of the nanosphere, the $y$ axis in Fig. 5 was normalized with respect to $V_S$. As with Fig. 2, the peaks found in the lower-energy region indicate the plasmon resonance, the positions of which are marked by yellow shaded stars, and the peaks of the Bennett state are indicated by the gray shaded stars. Figure 5 shows two characteristic tendencies for varying $a$. First, the maximum values of $S_{\text{abs}}$ were slightly decreased for larger values of $a$, whereas the spectral widths were broadened. The area of the peak that is proportional to the oscillator strength is found to decrease slightly (3-4%) for larger value of $a$. This phenomenon of slight decreasing and broadening may be attributable to the propagating effect as well as non-diple effect that are enhanced for larger nanospheres. Secondly, the spectrum of the plasmon resonance (Bennett state) was found to be blueshifted (redshifted) for larger values of $a$. This can be attributed to the volume ratio between the uniform electron density inside the nanosphere and the electron spill-out near its interface. Because the length of the electron spill-out has been fixed by setting $\lambda = 1/2$, the nanosphere associated with larger values of $a$ becomes similar to the characteristics defined by the hard-wall boundary condition applied in classical electromagnetism, since the bulk effect eventually drives out the surface effect. These changes with respect to the linear optical response have already been discussed in previous studies on QHT [23, 25].

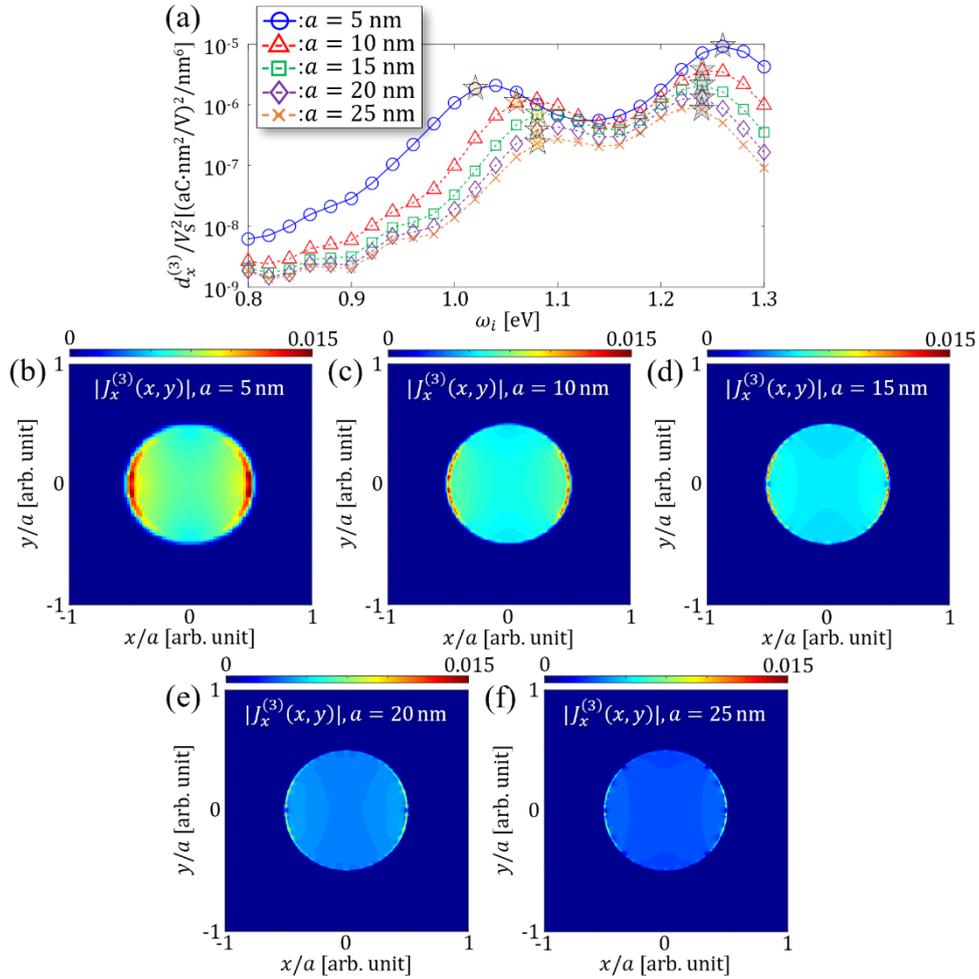

FIG. 6. (a) $\omega_i$ dependence of the third-order optical nonlinearity $d_x^{(3)}$, as defined by Eq. (22), for nanospheres with $\lambda = 1/2$ and $a = 5$ (solid blue line with circles), $10$ (dashed red line with triangles), $15$ (dashed green line with squares), $20$ (dashed purple line with diamonds), and $25$ (dashed orange line with crosses) nm. The vertical axis has been normalized with respect to $V_S^2$ because $d_x^{(3)}$ is proportional to it. The yellow shaded stars indicate the $\omega_i$ values that approximately satisfy one-third of the plasmon resonance energy, whereas the gray shaded stars denote the Bennett state. (b)–(f) Spatial distributions of $|J_x^{(3)}(x,y)|$ (see Eq. (21)) for each $a$ value condition; the $\omega_i$ values applied to obtain the results are denoted by the yellow shaded stars in (a). The $x$ and $y$ axes have been normalized with respect to $a$, but the color scale has not been normalized.

Lastly, as with Fig. 4, $d_x^{(3)}$ was calculated using Eq. (22), which reflected the third-order optical nonlinearity produced by an incident pulse with a fundamental frequency $\omega_i$; the other parameters related to the pulse (see Eq. (18)) were set to be the same as those used in the previous subsection. Figure 6(a) shows the $d_x^{(3)}$ results obtained under the conditions of $a = 5, \ 10, \ 15, \ 20,$ and $25$ nm. As described by Eq. (22), $d_x^{(3)}$ was derived from the square of $d_x$, and $d_x$ is also proportional to $N_e = n_s V_S$; thus, the vertical axis was normalized with respect to $V_s^2$. As before, the yellow shaded stars in the figure denote the $\omega_i$ values that approximately satisfy one-third of the plasmon resonance energy, whereas the gray shaded stars denote the Bennett state. As was observed in Fig. 4, specifically the results obtained under the conditions of $\lambda = 1/2$ and $a = 4.3$ nm, the $d_x^{(3)}$ results shown in Fig. 6(a) exhibited two peaks that reflect the plasmon resonance and Bennett state, which were respectively blueshifted and redshifted as $a$ increased. The maximum values for the linear optical absorption $S_{\text{abs}}$ in Fig. 5 were nearly the same following normalization. However, the third-order nonlinearity ($d_x^{(3)}$) results presented in Fig. 6(a) show that different values of $a$ corresponded to different maximum values even after the normalization. For example, in Fig. 6(a), $d_x^{(3)}$ calculated from $a = 5$ nm was nearly 10 times larger than that for $a = 25$ nm, indicating that the third-order nonlinearity per unit volume was enhanced by decreasing $a$. This finding has very intriguing implications for metamaterials, periodically arrayed nanostructures. Particularly, metamaterials composed of nanospheres with $a = 5$ nm or $a = 25$ nm that also maintain the volume ratio between the nanosphere and unit cell would have nearly identical linear optical absorptions, although they would likely have a significantly different third-order nonlinearity.

To explore the mechanism that determines the $d_x^{(3)}$ for each $a$, $|J_x^{(3)}(x,y)|$, as defined in Eq. (21), was again evaluated. Figure 6(b)–(f) display the results for each $a$; the applied $\omega_i$ values correspond to the values marked by yellow shaded stars in Fig. 6(a). The $x$ and $y$ axes were normalized with respect to $a$, but the color scale has no normalization. Figure 6(b), which shows the results for $a = 5$ nm, can be observed to have a spatial distribution that is very similar to that shown in Fig. 3(c), in which the nonlinear current density at the electron spill-out near the edges of the nanosphere is more enhanced than that of inner region. In the cases of the results shown in Figs. 6(c)–(f), however, the volumes of such electron spill-out component were relatively small, and considerably decayed nonlinear current was present throughout the interior of the nanosphere. Thus, Fig. 6(a) demonstrates that the third-order optical nonlinearity per volume tends to significantly decrease with increasing values of $a$.

**IV. CONCLUSION**

In this study, the third-order optical nonlinearity of metallic nanospheres was investigated by performing QHT-TF$\lambda$vW-based calculations. In particular, how electron spill-out at the surface of a nanosphere contributes to the third-order nonlinearity has been clarified. As the first step, the length of the electron spill-out was adjusted by varying the phenomenological parameter $\lambda$ for a small nanosphere with a diameter of 4.3 nm. Then, the third-order optical nonlinearity was strongly enhanced by increasing the amount of electron spill-out. In the second step, the size dependence of the nanosphere was investigated with respect to the third-order optical nonlinearity and under the condition of a fixed electron spill-out length. The diameter of the nanospheres was varied between 5 and 25 nm. The results revealed that, intriguingly, the third-order optical nonlinearity per volume was enhanced as a result of decreasing the diameter, whereas the linear optical absorption per volume was nearly unchanged. The electron spill-out played a key role in size dependence, as the surface-to-volume ratio was found to determine the third-order optical nonlinearity. These findings provide novel insight that is believed to be essential to the realization of metamaterials with large nonlinearity. Furthermore, the results presented here are also believed to provide useful information for researchers in the fields of nonlinear plasmonics and quantum mechanical physics.

**ACKNOWLEDGEMENTS**

This research was supported by the JST–CREST under Grant No. JP–MJCR16N5, the JSPS Research Fellowships for Young Scientists, and the JSPS KAKENHI under Grant Nos. 20J00449 and 20H02649. Calculations were carried out at Oakforest–PACS at JCAHPC through the Multidisciplinary Cooperative Research Program at the Center for Computational Sciences, University of Tsukuba, and at the supercomputer Fugaku through the HPCI System Research Project (Project ID: hp210137).